\documentstyle[12pt]{article}
\input epsfig
\begin{document}

\title{Dissipative motion in galaxies with non-axisymmetric potentials}
\author{Amr  El-Zant\\
{\em Physics department}\\
{\em Technion --- Israel Institute of Technology}\\
{\em Haifa 32 000, Israel}\\}

\date{ }
\maketitle

\abstract
In contrast to the case with most other topics treated in these
proceedings, it is not clear whether galactic gas dynamics can be discussed in 
terms of standard hydrodynamics. Nevertheless, it is clear that certain generic 
properties related to orbital structure in a given potential  and the effect of
dissipation can be used to qualitatively understand gas motion in 
galaxies.
The effect of dissipation is examined in triaxial galaxy potentials with 
and without rotating time dependent components. 
In the former case, dissipative trajectories settle around closed loop 
orbits when these exist. When they do not, e.g.,  
inside a constant density core, then the only attractor is the centre and this 
leads to mass inflow. This provides a self regulating mechanism for accession
of material towards the centre --- since the formation of a central masses 
destroys the central density core and eventually stops the accession. 
In the case when a rotating bar is present, there are usually several types of
attractors, including those on which long lived chaotic motion can occur
(strange attractors). Motion on these is erratic with large radial and vertical
oscillations.

\vspace{0.5cm}

\noindent \leftline{PACS: 98.10.+z,95.10.Fh,98.62.Gq} 
\leftline{Keywords: galaxies: evolution; ISM:general;galaxies: kinematics and dynamics}

\section{Assumptions of standard galactic dynamics}
\label{intro}

  It is usually assumed that test particles moving in a
galaxy are influenced by the mean field produced by all the material in the 
galaxy, discreteness effects being negligible. This is no trivial 
simplification: it reduces a problem with $3N$ coupled degrees of freedom to 
$N$, 3 degrees of freedom problems (with $N \sim 10^{10}-10^{11}$). 
It is also customary to assume that present day galaxies are in a steady 
state. Each object moving in the time independent  galactic potential
will thus conserve its total energy.  In addition, until recently,
 most systems studied had some ``special symmetries'', such that additional
quantities were conserved along the motion of test particles. For example, 
in spherical systems, the three components of angular momenta are also
integrals of motion, in axisymmetric systems, one of these is conserved.

 The structure of the  Hamiltonian equations, them being ``skew symmetric''
 --- symmetric with respect to the conjugate variables except for the minus 
sign in one of them --- requires that conserved quantities come in pairs.
Thus, by finding three constants of the motion, one secures all six constants 
required for solving the motion of a particle in three 
dimensions. Only two constants are required in two dimensions. 
If these were found by symmetry arguments, the system is said to be integrable. 
Such models of galaxies include those with ``trivial'' symmetries, such as 
spherically symmetric systems, or thin axi-symmetric ones
(where motion in the  vertical direction can be assumed to be de-coupled), 
as well as systems with more subtle symmetries such as the non-axisymmetric 
Stackel potentials (for an extensive  discussion of large classes  
of non-axisymmetric galaxy models see de-Zeeuw and Pfenniger 1988).

Integrable systems are characterised by the interesting property that the 
motion can be decoupled into independent one dimensional oscillations, once
a suitable coordinate system has been found. Thus, a three degrees of 
freedom problem is further  reduced to three one degree of freedom ones. 
This is
again not a trivial simplification. One has now reduced the a gravitational
$N$-body problem to that of studying a collection of independent oscillators. 
In practice, strictly speaking, no $N$-body ($N>2$)
 gravitational system can be exactly
integrable, since it was proved by Poincar\'e that one cannot find enough
constants of motion which exist for all initial conditions. Nevertheless, 
it is assumed that at least for times comparable to a Hubble time, where 
discreteness effects are small, some gravitational 
systems can be thus described. 
Another limitation is that completely 
integrable systems are rather rare (they actually form a set of measure zero
in the space of possible systems). Nevertheless, the celebrated KAM 
theorem (e.g., Arnold 1987)
 guarantees that as long as the considered systems are not too different 
from integrable ones, most initial conditions will lie on trajectories that are 
qualitatively similar to those of neighbouring completely integrable systems.
In the following section we discuss an example of a potential where
the above conditions hold for a large range of parameters and initial 
conditions.

\section{Of potentials and pendulums}

Observations indicate that, after an initial rise which is usually more or 
less linear, the circular velocity of test particles in galactic potentials
is constant over a large range of radii. Such a situation can be modelled by a 
particularly simple potential of the form

\begin{equation}
\Phi = \frac{1}{2} v_{0}^{2} \log(R_{0}^{2} + R^{2}),
\end{equation}
where $R_{0}$ is the ``core radius'', beyond which the rotation
curve flattens.
This potential can be generalised to simulate a triaxial figure and has been
studied in some detail in the book by Binney and Tremaine (1987) (BT).
In general, it  can be written in terms of Cartesian coordinates as 
\begin{equation}
\Phi_{H} = \frac{1}{2} v^{2}_{0} \log 
\left( R^{2}_{0}+x^{2} + p y^{2} + q z^{2} \right).
\end{equation}
The equipotential are ellipsoids with ratios $1/p^{2}$ and $1/q^{2}$
between the middle and long and short and long axis respectively.

Within the core radius, the density is nearly constant and solutions of 
Poisson's equation predict a nearly quadratic potential. In such {\em harmonic}
potentials the motion is separable in Cartesian coordinates. In the $x-y$
plane for example,  orbits can be represented as independent superpositions 
of oscillations
in the $x$ and $y$ directions. They are said to be parented by the $x$ and 
$y$ axial orbits which move up and down these axes. Such ``box orbits''
 have no definite sense of rotation: their net,  time averaged, angular 
momentum is zero. In the region outside the core radius other types of 
orbits may exist.  In particular, loop orbits, which do have definite sense 
of rotation, appear. These are parented by the closed loops, which are 
oval, non-self-intersecting closed periodic orbits.

Since the motion of most orbits in nearly integrable potentials can
be represented as a superposition of one dimensional oscillations, it is
instructive to invoke the analogy with this most familiar of 1-d 
oscillations: the simple pendulum. For small oscillations, the motion 
is dominated by the linear term in the force field and resembles that of 
a linear oscillator ---  where the period of oscillation is independent
of the amplitude. For larger amplitudes, the period of oscillation increases
with amplitude and becomes infinite at the separatrix. Beyond this, a
different type of qualitative motion appears. Now, instead of  
oscillating, orbits of the pendulum have a definite sense of rotation about
the centre.

In polar coordinates then, box orbits can be characterised as two independent 
librations, while loop orbits are similar to one librating motion 
and rotating one. Since the frequencies will, in general, be incommensurable,
most orbits of both types will not close. The exception being the periodic 
orbits. In the box family these resemble the familiar Lissageou figures, in
the loop family these are closed loops. 

Deep inside the harmonic core, the motion then resembles a superposition of 
two uncoupled harmonic oscillators. Beyond that,  it is possible 
to have rotation in one of the coordinates, and the oscillations change 
frequency with amplitude. It turns out however that the ratio of the rotation 
frequency to that of oscillation around the closed loop orbits is usually more
or less constant (BT), thus these continue to behave as if they are uncoupled
 and
loop orbits, when they exist, are usually stable. The situation with
the box orbits is different however. At large radii the oscillations become 
coupled and therefore the motion can no longer be represented as independent 
oscillations and no longer mimics the motion in integrable systems. This 
gives rise to ``chaotic'' trajectories which eventually populate most of the 
region once occupied by the box orbits (Schwarschild 1993).

The advent of chaos can also be understood with the help of the pendulum 
analogy (see, e.g., Zaslavsky et al. 1991 for a fuller account of the 
following discussion). 
Near the separatrix the quantity 
$\frac{1}{\omega} \frac {d \omega} {d A}$
 goes to infinity. That means that
any small change in the amplitude $A$ will lead to large changes in the 
frequency $\omega$ ---  which implies large changes in the phases of 
trajectories with nearby initial conditions differing slightly 
in amplitude. In addition, very small perturbations can transform 
oscillating trajectories into rotating ones (and vice versa), and also
cause rotating trajectories to change their sense of rotation.
Thus, introducing periodic perturbations, for example, can lead to extremely
complicated trajectories around the separatrix. These ``homoclitic tangles'' 
 are the types of trajectories Poincar\'e  thought were so complicated that
he did not even attempt to draw them. Schematic representation however
was achieved by Arnold and these could be found in many standard texts 
(e.g., Lichtenberg and Lieberman 1992 (LL)).
The larger the perturbation, the larger this  
``separatrix layer'' where the
chaotic motion described above takes place. 

In multidimensional systems,
the role of the  unstable equilibrium point around the separatrix is replaced by 
unstable periodic orbits. When these form a dense set, the chaotic regions
merge. In the process of this transition to chaotic behaviour two points are of 
importance: nonlinearity and symmetry. Without nonlinearity, the frequencies
are constant with amplitudes and so the ratios of frequencies of different 
degrees of freedom are likely to be irrational --- i.e., no periodic orbits
and no separatrix layers. As with the pendulum, equilibrium solutions are 
produced by a  break in symmetry: in the pendulum, when gravity is turned on, 
we get 
the stable and unstable equilibria at the bottom and top respectively.
The stable equilibrium parents the oscillations and the unstable one repels 
nearby trajectories.
Integrable systems have the exceptional properties of having only a finite 
number
of periodic orbits that are stable or unstable at a
given energy --- the rest being marginally stable (see BT). The stable periodic
orbits parent the regular general orbit families 
(e.g., the box and loop orbits) which occupy almost all of the phase space.
Lack of symmetry produces periodic orbits (originally in pairs of stable and 
unstable ones, but further increase in the perturbation leads to increase in
the number of unstable orbits). In this case there is a non-zero measure of 
chaotic orbits.
Finally, in a multidimensional system,
the ``perturbation'' that may give rise to chaotic orbits can either 
be external (as in a time dependent term in the potential) or coupling 
between different degrees of freedom. This coupling requires non-linearity 
and asymmetry.

\section{The effect of dissipation: attractors}
 
The skew symmetric form of the Hamiltonian equations ensures that 
any expansion of the phase flow in any direction will be counteracted 
by contraction in a conjugate one  (which thus ensures
conservation of phase space volume: for a discussion of the geometry of
this  {\em symplectic structure} see e.g., Arnold 1989; Marsden \& Ratiu 1994;
von Westenholtz 1978). 
Dissipative systems, 
which have velocity dependent force terms, on the other hand
  have a ``time arrow''.
The irreversible behaviour 
of dissipative systems  is manifested in the contraction 
of the phase space volume (corresponding to a set of initial conditions 
and evolved dynamically). Trajectories thus end in ``attractors'', which in
general will have dimension less than that of the embedding phase space
(for an excellent review  of concepts related to the behaviour of dissipative 
dynamical systems see  Eckmann and Ruelle 1985; 
collections of original articles can be found in Hao 1989).

 Any given system can have more than one attractor. On which attractor a 
given initial condition will end up will depend on which basin of attraction
it started from. Asymptotic dissipative motion in generic galactic mass 
distributions is expected to exhibit more than one attractor. Here again
the analogy with  pendulae is helpful. Intuitively, one can see that 
friction would force a system of rotating pendulae to move at the same 
rotation speed --- or if this is not possible, then at least with the least
relative velocity (so as to be compatible with the least possible dissipation).
Thus, the symmetry of the problem is such that 
  long lived states where there is common 
rotation can persist. Translated into the language of galactic dynamics, this
implies that a collection of loop orbits interacting in a dissipative manner can 
keep their definite sense of rotation (assuming that the dissipation mechanism 
does not lead to net loss in the  total orbit averaged angular momentum).
The oscillatory motion however would have to die out --- again in  a system of 
pendulae oscillating around the stable equilibrium, dissipation, by symmetry,
would have to lead to all pendulae ending up at rest at the stable equilibrium.

The above argument implies that in a non-axisymmetric galactic potential,
the fate of dissipative trajectories will depend on their type. Loop orbits 
would end up following up the closed periodic loops --- since the radial 
oscillations would die out while net rotation can persist --- while box
orbits would dissipate towards the centre (because oscillations in both
coordinates would die out). The former being long lived {\em limit cycles}
while the later are attracting {\em fixed points}.
The situation in the chaotic region of the original Hamiltonian system 
will depend on whether there is any energy input. If the Hamiltonian 
decreases constantly, then the dissipative orbit passes by successive 
chaotic and regular regions before finally dissipating towards the centre
--- such behaviour is sometimes labelled ``transient chaos'' (see, e.g., LL).
If, on the other hand, there is some forcing in such a way that the Hamiltonian
does not not decrease monotonically, then some trajectories may end up 
on {\em strange attractors} on which long lived chaotic motion can occur
(for more details and references on this intricate subject see 
Pfenniger and Norman 1990 (PN)).

\section{Modelling dissipation}
\label{modis}

The motion of stars in galaxies is essentially collisionless. Effects 
due to dissipation are therefore expected to be important only for the 
gaseous component, which plays an important role in the evolution of
(especially disk) galaxies.
One important characteristic that has to be taken into account when attempting
to model the interstellar medium is its highly clumpy,
non-uniform nature. This property renders standard hydrodynamical treatments
based on the continuum approximation and an equation of state not including 
gravitational effects inadequate,
since the assumption of local thermodynamic equilibrium stemming from 
a certain separation of ``fast'' and ``slow'' processes 
is no longer satisfied in a straightforward manner. Standard thermal
physics and the accompanying hydrodynamic equations therefore no longer
apply (Pfenniger 1998) --- since things like the Chapman-Enskogg approximation
(Andersen 1966) are no longer valid. Thus, in this paper, as opposed to most presentations
in these proceeding, ``fluid dynamics'' is not synonymous with standard 
hydrodynamics. This is true of systems whose character is determined by
 gravitational instability, which guarantees that they behave like systems
near a phase transition, which in turn lack characteristic scale. 
The lack of such a 
characteristic separation of space and time scales renders standard 
hydrodynamics inaccurate (e.g., Lebowitz et al. 1988).
The clumpy and apparently scale
 invariant nature of the gaseous interstellar
medium has  inspired attempts to treat it as a fractal object (Pfenniger
\& Combes 1994). This method still awaits detailed application to realistic 
situations.

Nevertheless, there are indeed fast and slow processes which 
can be separated. These are the collision time between 
gas clouds, which is of order of one
to ten Myr, and the dynamical time which is much longer --- being of the
order of  60 to 600 Myr. 
This leads to a formulation where  one considers the hydrodynamics of 
collections of gas clouds which interact with each other over time-scales
which are short compared to the dynamical time 
(details of the grounds on which such an approach may be justified are given 
by Scalo \& Struck-Marcel 1984).
 A new set of hydrodynamic
equations,  more appropriate to this situation, is thus obtained. In these
equations a ``fluid element'' is therefore a region small enough so that
the macroscopic gravitational field can be considered  roughly constant  
while large enough to contain a fair sample of gas clouds. According to
Combes (1991) the total mass of molecular Hydrogen in the Milky Way for
example is about $2-3 \times 10^{9}$ solar masses concentrated in clouds
of  mass greater than $10^{5}$ solar masses.   At about 10 kpc
these are concentrated in a region of a hundred pc from the plane of the
disk. Therefore the volume mentioned above would be of the order, say,
500 by 500 by 100 pc, but of course will be smaller in the central areas
where the concentration of molecular hydrogen increases significantly 
(Combes 1991). 

Assuming that the hydrodynamic effects 
(e.g., pressure, viscosity etc.) are
small (i.e., gas clouds move primarily under gravitational forces)
and that any evolutionary effects are slow one can imagine 
the full hydrodynamic
equations of these macroscopic elements
to be perturbations to the collisionless Boltzmann equation. 
 Here we will follow
Pfenniger \& Norman (1990) (PN) and assume that in the context discussed above,
the  main hydrodynamic effect can be approximated to first order as a velocity
dependent viscous force, the value of which is   small in comparison to 
the value of the mean gravitational field. In PN, dissipative perturbations
of the form
\begin{equation}
F_{fric}= - \gamma v {\bf v},
\label{tax:difor}
\end{equation}
where  $\gamma$ is a constant which determines the strength of the dissipation,
were used. 
This form is similar to the one obtained if one  assumes that the 
individual clouds are composed of infinitely compressible isothermal
gas. Then in each collision (Quinn 1991)  the acceleration
during an encounter of two clouds is given by 
\begin{equation}
a_{en} \sim u^{2}/r,
\end{equation}
where $u$ is the absolute value of the relative velocity of the clouds and
$r$ is their separation. 

One can to first order estimate $\gamma$ using the usual ``mean free path''
approximation  of (linear) viscosity:
\begin{equation}
\gamma_l = n \lambda v_{rms},
\end{equation} 
where $n$ is the volume number density of gas clouds, $\lambda$ the mean free path and
$v_{rms}$ is the average random velocity. Supposing that the volume density is
of the order of a few hundred or so per cubic kpc and the random velocity of the order
of a few parts per hundred kpc/Myr, so that the mean free path is also of that
order if the mean free time is about 1-10 Myr. 
This gives a value $\gamma_l = \gamma \times v_{rms} \sim 0.01-0.1$ . 

 PN   used  even  more conservative values
for $\gamma$ and still found  highly non-trivial effects. It was
shown in the aforementioned paper that even extremely small dissipation rates can be
greatly amplified in the presence of resonances.  The main resonance existing in the 
 disk-halo systems is the $1:1$ resonance.
This is the ``separatrix'' between the area where loop orbits occur and the 
nearly harmonic core. This resonance  is not accompanied by widespread
instability because the closed orbits that bifurcate from it are the stable closed
$1:1$ loops. Nevertheless, the effect is still palpable, for even in completely integrable
one-dimensional systems, separatrix crossing can lead to faster dissipation 
(Parson 1986).
Dissipation is basically a product of disorganised motion, 
in that context either chaotic
resonance layers or the effect of the abrupt changing of qualitative motion at a separatrix
can cause acceleration of dissipation even for systems as simple as 
pendulae (PN).
In particular bifurcations leading to the destruction of stable limit cycles where
dissipative orbits can settle are particularly effective as we shall see below.

\section{\bf Dissipation in non-rotating systems}
\label{tax:dispdisk}

Using a value of $\gamma=0.005$ we have integrated the equations of motion 
for particles moving in the  logarithmic potential
and being influenced by dissipative  forces of the form 
given by Eq. (\ref{tax:difor}).  
The parameters for the 
logarithmic potential are taken as $R_0=6$ kpc and $v_0=0.19$ kpc/Myr,
with axis ratios $b/a=0.9$ and $c/a=0.8$.

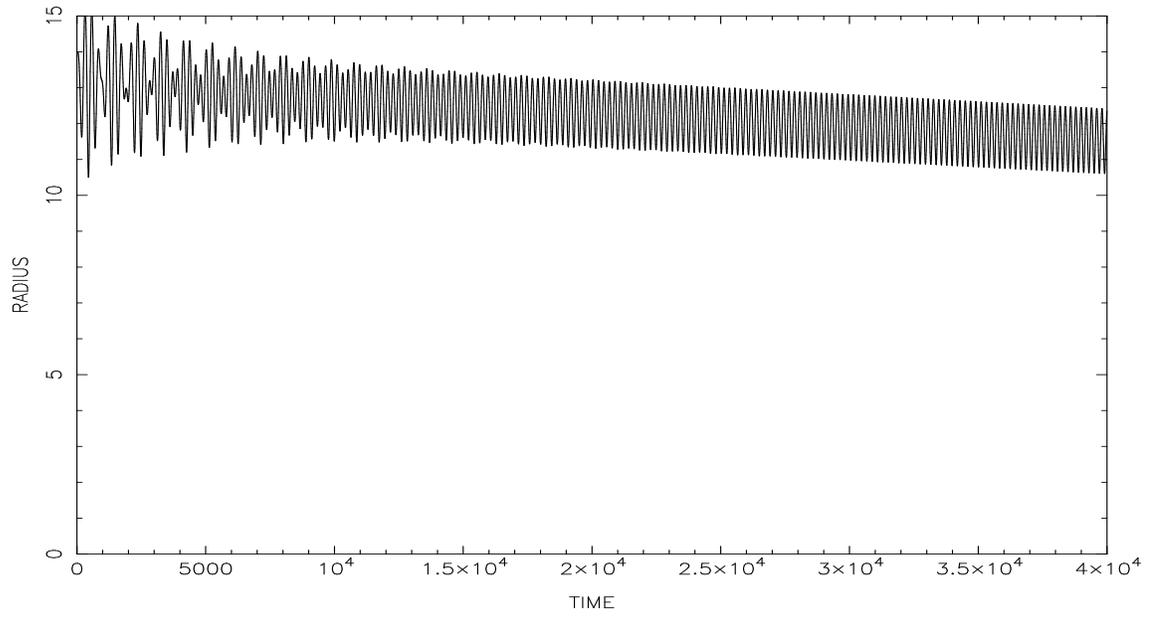
\begin{figure}
\caption{\label{dispq}
 Evolution of radial coordinate of trajectory starting from an initial value of 14 kpc}
\end{figure}

The quantity ${\bf v}$ in Eq.~(\ref{tax:difor}) is taken to be the 
vector difference 
between the velocity of the particle and that of a particle moving with the
local circular velocity.  Such motion cannot exist, because in a
non-axisymmetric potential the orbits with smaller radial departure from
the mean are ovally distorted closed loop orbits. Because the dissipation 
rate is small however, far from the core, where these orbits are fairly
close to circular, the difference in velocities is fairly small and an
orbit oscillates around the closed loops for very long times (Fig.~\ref{dispq}). 
This means that effectively our dissipative point particles representing
fluid elements of gaseous disks
settle into
regular quasiperiodic limit cycles.  In the case where  we would calculated
$\bf v$ in terms of the difference between a particle's trajectory and the 
local periodic loop orbit we would obtain a true limit cycle as the
asymptotic {\em attractor}.

The situation is rather different however as one moves nearer the 
halo core. In this case,
the closed loop orbits become more and more eccentric and cannot parent any
orbits that spend all their time inside the core. One can then 
say, for trajectories inside the core, the only attractor is the centre. That
is, these trajectories are outside the basin of attraction of limit cycles
represented by the closed loops.  

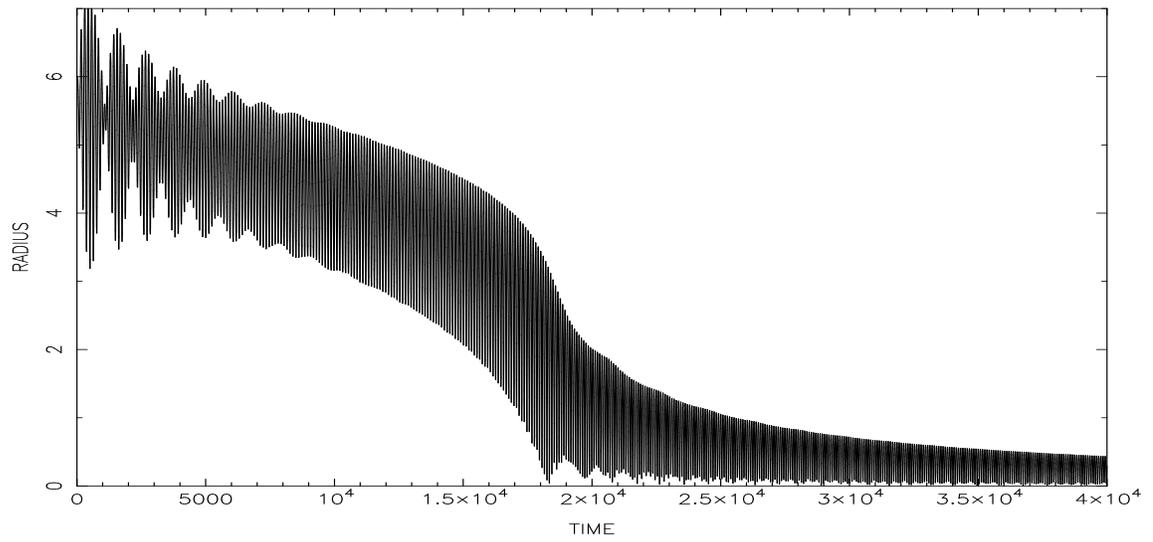
\begin{figure}
\caption{\label{dispa}
 Evolution of radial coordinate of trajectory starting from an initial value of 6 kpc}
\end{figure}

In Fig.~\ref{dispa} we plot the radial coordinate of a trajectory
of a  particle which starts on the $x$ axis at
a distance of $6$ kpc. As can be seen, as one moves closer and closer towards 
the centre,
the dispersion in the radial coordinate of the particle increases as the local
loop orbits become more and more eccentric. At the separatrix 
 beyond which
no  loop general loop orbits can exist (which occurs at about 4 kpc)
there is a sharp transition in the dissipation rate and the particle quickly moves 
towards the centre.

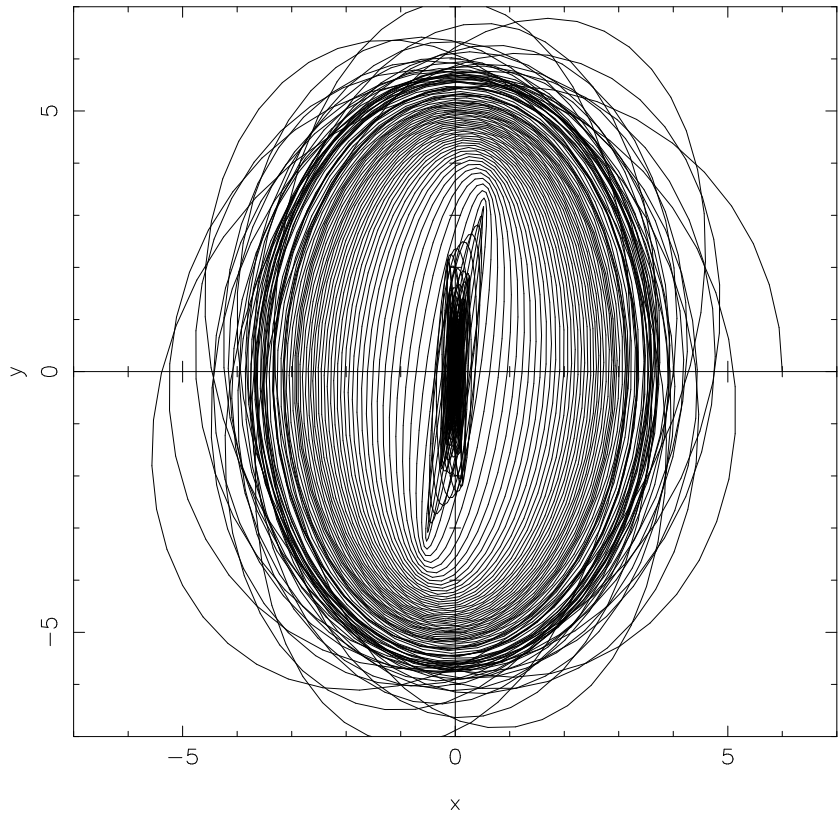
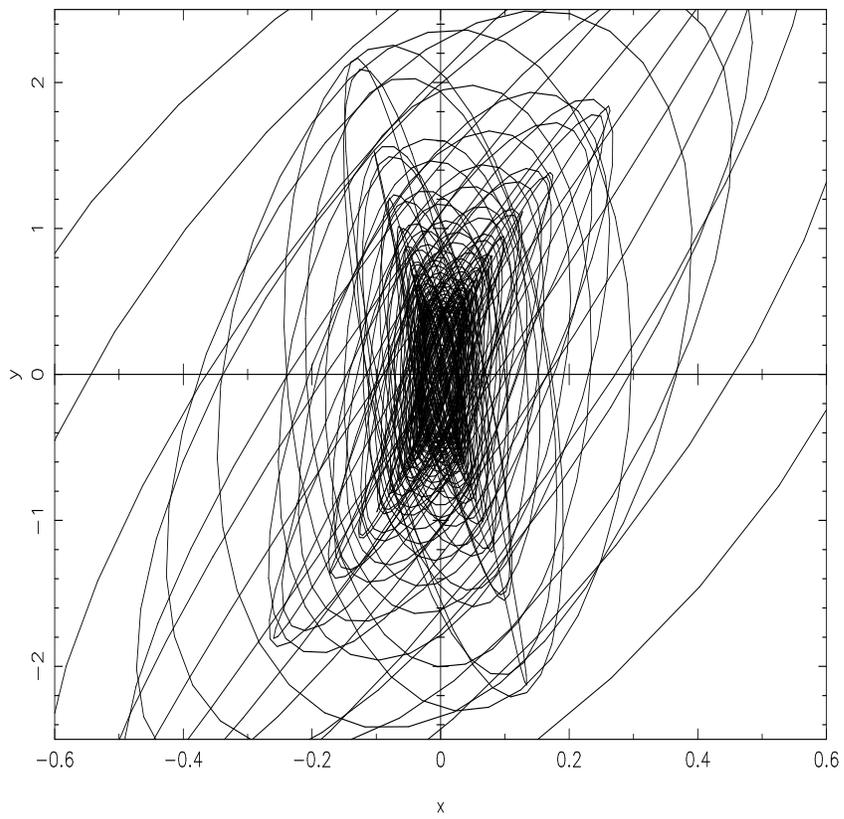
\begin{figure}
\caption{ \label{dispaxy}
 Trajectory in the $x-y$ plane of orbit with radial coordinate represented in Fig.~\protect\ref{dispa}
and a more detailed view of the segment of the trajectory from 1750 Myr 
to end of the run at 50 000 Myr (bottom)}
\end{figure}

Fig.~\ref{dispaxy} displays the spatial evolution of this orbit,
 which shows it to follow a
sequence of loop orbits followed by what essentially are sections of box orbits.
This latter behaviour, which starts roughly at about 16000 Myr and increases
the dissipation rate dramatically, is shown in the plot on the right hand side
of this figure. 
This process of course will lead to the growth of central mass concentrations.
However it appears that this process slows down significantly as the central mass
increases. The central mass ruins the harmonic nature of the potential;
the $1:1$ resonance is brought inwards nearer to the
centre of the potential. In addition the potential
of course becomes much more symmetric in the inner regions so that the closed
loop orbits, which now exist in the central areas, 
are close to circular.

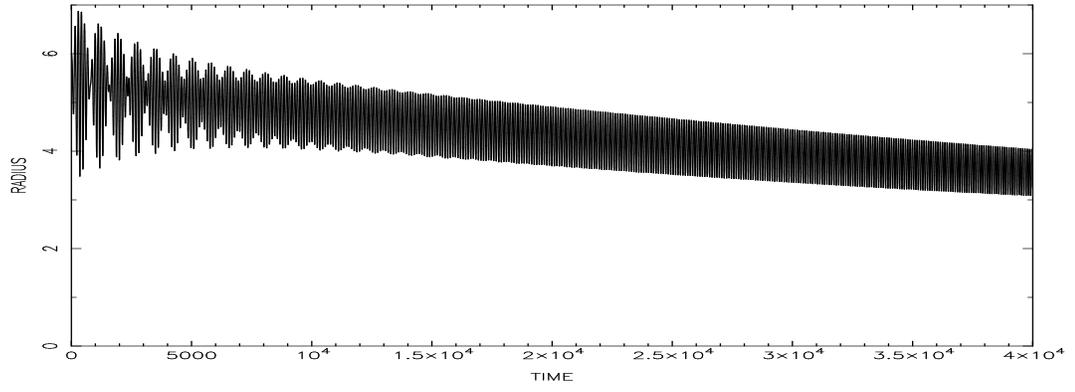
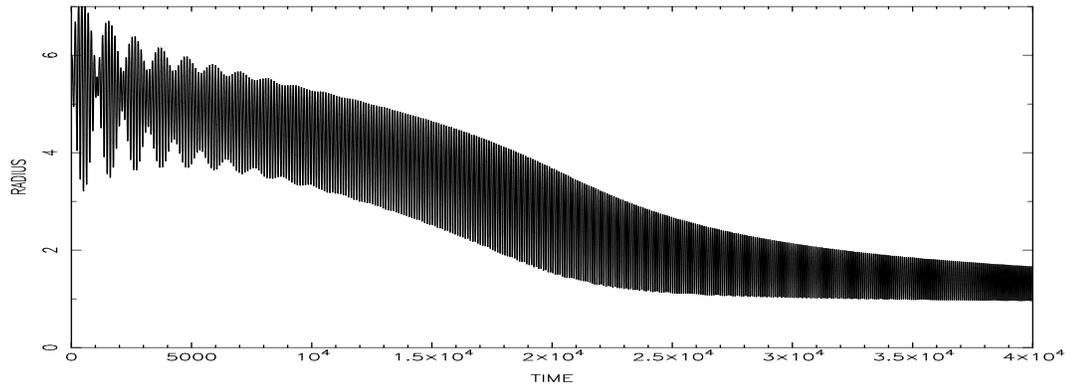
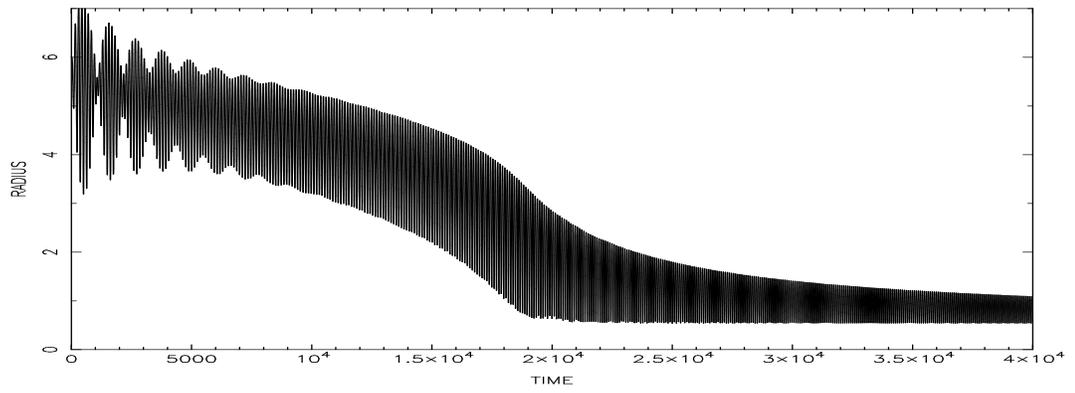
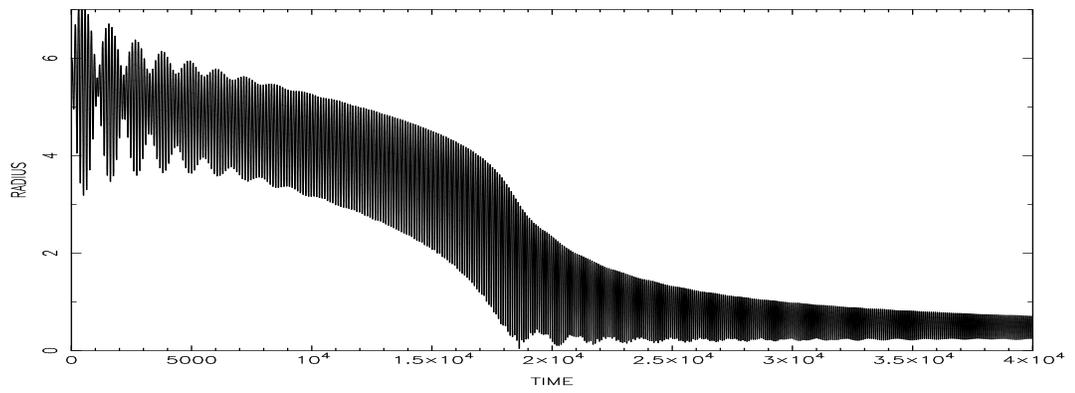
\begin{figure}
\caption{\label{dispc}
 Same as in Fig.~\protect\ref{dispa} but when a central mass of (from top to bottom)
$GM_{C}=0.01$, $GM_{C}=0.0005$, $GM_{C}=0.0001$, $GM_{C}=0.00001$}
\end{figure}

The top diagram in Fig.~\ref{dispc} shows the behaviour 
starting from the same initial conditions as in
Fig.~\ref{dispa} and exactly the same potential except that a central mass of 
$GM_{c}=0.01$
is present (in the units we are using $G=1$, distance is measured in kpc and
time in Myr, so that $GM=1 \sim 2 \times 10^{11}$ solar masses). 
Clearly the behaviour previously observed --- that is the accelerated 
dissipation rate --- is no longer present. Smaller central masses lead to more
eccentric loop orbits. However, even for very small central masses (anything
greater than $GM_{C} \sim 0.00001$) loop orbits exist deep inside the core. 
Subsequent plots of Fig.~\ref{dispc} 
clearly show that there is a certain region between the halo core radius and 
the centre
where the trajectories are extremely 
eccentric and the dissipation is very large. 
The scope of this region increases with decreasing
value of the central mass.  If this central mass is large enough, then the whole
core region would support stable loop orbits. Otherwise there will be an annulus
where no stable non self-intersecting periodic orbits exist. In the Hamiltonian 
limit, this region will contain chaotic orbits and higher order box  orbits.
Dissipative trajectories will therefore rapidly spiral towards the centre, until 
the region where loop orbits exist is reached.  
As they move through the chaotic region,
it is also possible for dissipative trajectories  
to be $z$ unstable given the right kind
of circumstances.
This behaviour has been observed for small $\gamma$ (about 0.0005)
and when the central mass was also small.
In this case,
the a dissipative trajectory visits the $z$ instability strips in the ``chaotic sea'' and 
does so slowly enough so as to acquire a large $z$ excursion, before reaching 
the stable limit cycle around the closed loops in the centre. 
Although this final motion is regular, any stars born in the intermediate process 
will have  chaotic trajectories with large vertical excursions (which will be enhanced
once dissipation ceases).

\section{\bf Dissipation in systems with rotating bars}

Our dissipative force tends to circularise the motion. In the
case when no bar was present this meant that general orbits would  
oscillate around the
stable loop orbits, unless none were to be found in which case they dissipated
 rapidly towards 
the centre. In the cases discussed here however such orbits can be unstable,
especially around the various axisymmetric resonances. 
For while for a nonrotating system 
these are not important 
(since ratio of the rotation frequency to that of small 
perturbation around nearly circular orbits is nearly constant), 
this is generally not the case in  a rotating system 
(where for a  general pattern
speed $\Omega_P$ and rotation frequency $\Omega$,
the ratio of the small perturbation to $\Omega - \Omega_P$ can vary
significantly with radius). The addition of a central mass increases this effect.

A general dissipative trajectory starting from the basin of attraction of the centre
will pass through the various vertical and horizontal  instability strips as it
moves inwards. The erratic flow across the resonances will lead to large oscillations in
both the $R$ and $z$ cylindrical coordinates. This will in turn lead to two effects
(discussed in detail in PN). The first will be again an increased dissipation rate and
the consequent decay in the $R$ coordinate. The second effect will be the scattering of
particle trajectories out of the disk plane, as one passes through the vertical resonances. 
 For certain dissipation laws, and especially if the potential is time dependent, 
it is also possible to find strange attractors, in addition to fixed points
 (like the centre)
and limit cycles, as the invariant limit sets of dissipative trajectories. In this case, 
the motion 
is chaotic for infinite times and not just when passing through resonances (this 
latter situation, which is much less well defined than the case when a strange 
attractor exists and which is similar to the situation described near the end 
of the previous section, 
is sometimes labelled `transient
chaos' or `intermittent chaos': see, e.g., LL for a review).   
We have not attempted here an extensive search of the phase space for all possible limit
sets. Nevertheless, a fairly large number of initial conditions and parameters
(chosen more or less
by random trials) were tested. In the following we describe a few (we hope representative)
examples.
The trajectories described here move in galaxy models with the same  logarithmic 
potential   parameters
 used in the previous section. A Miyamoto Nagai disk with potential
\begin{equation}
\Phi_{D}= -  \frac{GM_{D}} {\sqrt{ x^{2}+y^{2}+ 
\left( a_{d}+\sqrt{b_{d}^{2}+z^{2}} \right)^{2}} },
\label{tax:disk}
\end{equation}
with $a=3$ kpc, $b=1$ kpc and $GM_D=0.3$, represents the bulge-disk contribution
(in the units we are using the gravitational constant is $G=1$ and time is 
measures in Myr which makes $GM=1$ equivalent to about $2 \times 10^{11}$ 
solar masses).
 A second order Ferrers bar (BT; Pfenniger 1984) with axes of 6 kpc, 1.5 kpc 
and 0.6 kpc 
and mass $GM_B=0.2$ is also added. The systems studied below will therefore represent disk
galaxies with stellar bars and triaxial dark matter haloes. The pattern speed
$\Omega_P$ of the bar is chosen so that corotation in the disk halo potential 
is at 6 kpc (i.e., the bar ends at corotation, as is believed to be the case in
most systems: Sellwood and Wilkinson 1993).

In the axisymmetric halo case, the triaxial perturbation is small. 
Near the end of the bar, closed loop orbits, which are termed $x_1$ in 
rotating systems, exist (see, e.g., 
Sellwood \& Wilkinson 1993 for a review of the orbital structure in barred potentials).
A trajectory starting with the circular velocity in the azimuthally averaged
potential ends up closely following one of the periodic $x_{1}$ orbits 
 elongated 
along the bar. This situation is analogous to the loop orbits
outside the core radius of a non-rotating potential (except that in the latter
case, loop orbits are elongated normal to the elongation of the mass distribution).
Some trajectories may alternatively end up
librating about one of the Lagrangian points  
near the end of the bar.  As one moves towards the centre of the potential
however, the gaps in the $x_{1}$ family are affected by lower order resonances.
The strongest of these is fourth order ``ultra-harmonic resonance''.
As trajectories pass this resonance they experience rapid decay in their radial
coordinate  --- an effect that is by now familiar.  There is also some
scattering in the $z$ direction for some initial conditions.
However this is not very dramatic (a few hundred pc) for the parameters chosen here. 
Since the potential  does not contain any other lower order resonances
near the centre, the dissipation rate therefore slows down considerably and 
the trajectory settles down into a quasi-steady limit cycle around the $x_1$ orbits.
Here, because the halo spherical, these may exist near the centre.

The addition of even a small central mass broadens the existing resonances and moves 
them outwards, and creates new lower order ones near the centre (PN).
Thus, while in the case of a non-rotating potential the loop orbits were stabilised 
by the addition of a central mass, here we have the opposite. The central mass creates
and broadens the axi-symmetric resonances and destroys the loop orbits. This is due to the
 effect described in the opening paragraph of this section.
  The effect of dissipation  therefore  becomes much more dramatic, 
with an attracting point {\em at the centre} quickly reached.
The scattering in the vertical direction is also enhanced. It is however
still comparatively small, and becomes significant (about a kpc or so) only
when the dissipation rate is small ($\gamma \sim 0.0002$). These values
are comparable to what was obtained by PN but are far smaller than in the Hamiltonian
case or when a triaxial halo is present.

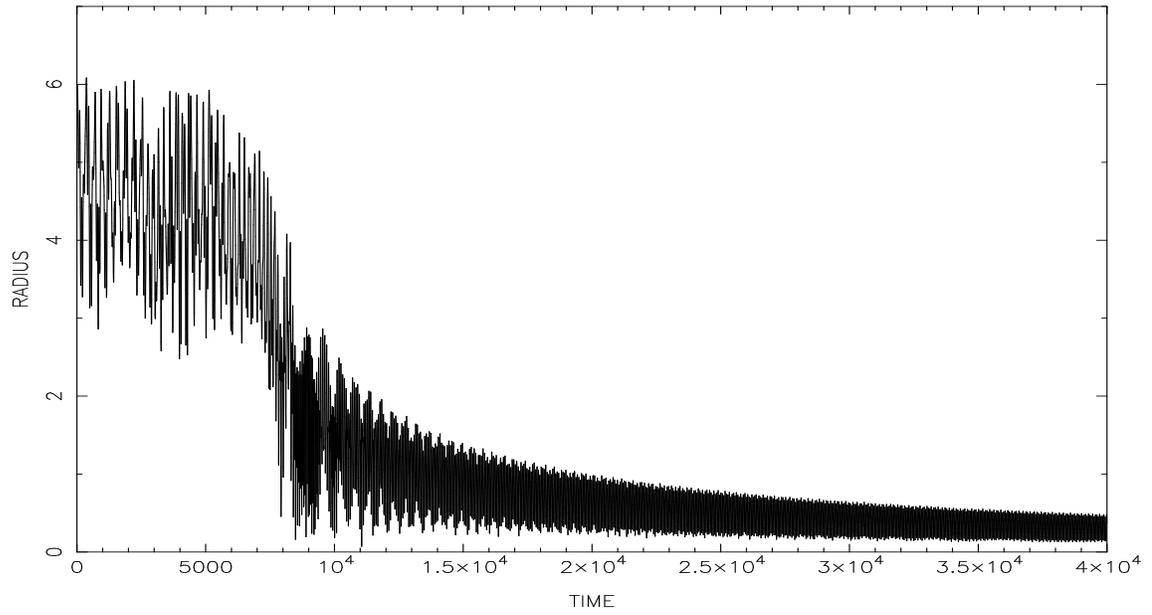
\begin{figure}
\caption{\label{dispr}
 Trajectory started from edge of a rotating  bar in logarithmic potential with axis rations $b/a=0.8$
and $c/a=0.7$}
\end{figure}

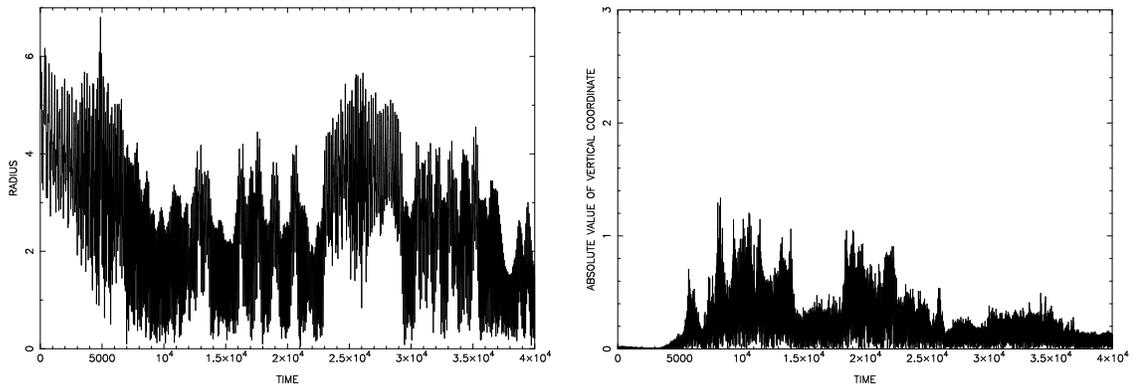
\begin{figure}
\caption{ \label{dispw}
Left: Same as in Fig.~\protect\ref{dispr} but with $\gamma=0.0005$. 
Right: Corresponding evolution 
of the absolute value of the $z$ coordinate}
\end{figure}

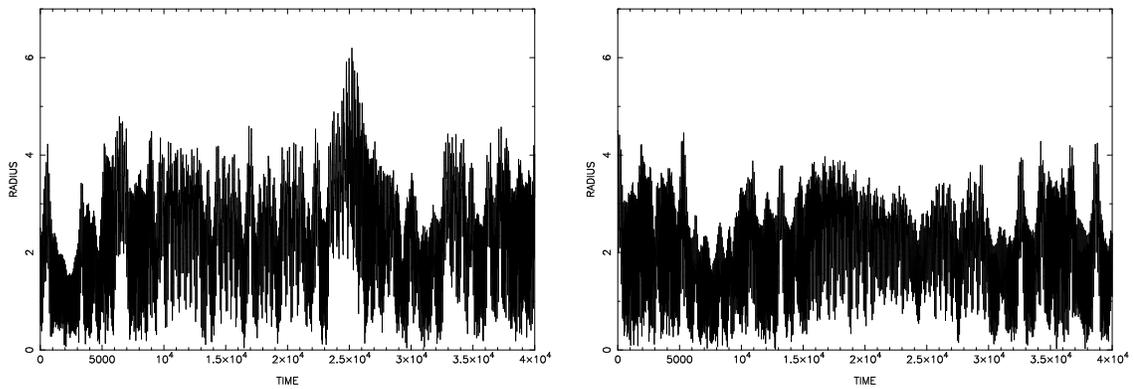
\begin{figure}
\caption{ \label{dispu}
 Evolution of radial  coordinates for trajectories starting at a 
only 2.5 kpc (left) and 4.5 kpc from centre of the mass distribution but appearing
 to end up in the same attractor as the trajectory in Fig.~\protect\ref{dispw}.
Here also $\gamma=0.0005$}
\end{figure}

We now turn our attention to the case where the halo is non-axisymmetric. 
 We fix the halo potential axis ratio 
in the disk plane to be $b/a=0.8$ while normal to the plane of the
disk we take  the axis ratio to be $c/a=0.7$. Fig.~\ref{dispr} shows the time evolution
of the radial coordinate of a particle  started from  the edge of the bar with
the velocity of a circular orbit in the azimuthally averaged potential.
As can be seen from that figure, even trajectories 
starting  this far out from the centre of the
potential can now be transported there. It can be noted however that after the
initial rapid decay in the radial coordinate, the trajectory settles down 
to a state where this coordinate oscillates about a value of roughly 0.5 kpc,
  with little
further decay in what appears to a be a regular limit cycle. Trajectories started
near the ultra-harmonic resonance are not attracted further in but are
caught in what appears to be either a limit cycle 
a strange attractor. Distinguishing between the two cases would require 
testing with a suitable indicator of chaos (e.g.,  
calculation of Liapunov exponents), in this case however
 the distinction is not really 
crucial practical importance, since the behaviour is similar. However,
when the dissipation was decreased  (by lowering the value of  $\gamma$
to $\gamma=0.0005$), it was found that some trajectories end up in states that
can be seen to be  highly erratic even by simple inspection. Fig.~\ref{dispw} displays
the radial coordinate time series of such an orbit. This behaviour is also usually
accompanied by non-negligible excursions in the vertical direction.
These are suppressed by high dissipation rates. It is also interesting to note that 
orbits starting from as small an initial radial coordinate as 2.5 kpc 
where particles are transported towards the outside and all
the way to the end of the bar appear to end in this attractor at least as far as
the radial coordinate behaviour is concerned (Fig.~\ref{dispu}).

In the presence of a stronger bar and in the absence of dissipation
(e.g., $GM_{B}=0.04$), many of the 
trajectories starting near the end of bar escape in the Hamiltonian limit. 
In the presence of significant dissipation however most of these trajectories end
up in stable attractors outside the bar. In general the 
motion may be extremely complicated with trajectories violently oscillating in and out 
through the disk plane as well as normal to it. 
This is especially true if the dissipation is weak.

\section{Concluding remarks}

The object of this paper was to present simplified models of the effects of 
dissipation in galaxies with non-axisymmetric potentials.
In a way, the study 
of this type of motion is much simpler than that of the Hamiltonian limit, since the
phase space collapses into a few asymptotic attractors along with their attracting basins, 
as opposed to the complicated `mixed' phase space of low dimensional Hamiltonian systems.

The approach used here boils down to adding weak (compared to the gravitational 
field)  dissipative perturbations to trajectories in non-axisymmetric galaxy models.
This approach avoids any detailed assumptions on the details of the dissipation process
and reveals generic features resulting from the structure  of the phase space 
{\em independent} of the dissipation law used (this has been checked). This 
avoids such questions as to what type of detailed  thermal physics  should be used in far
from equilibrium systems such a the clumpy structures of the inter-stellar medium 
where conventional thermodynamics fails
(further discussion of this point can be found in Section~\ref{modis}; 
see also the interesting
discussion on this subject given by Pfenniger 1998).

In the case when no rotating bar is present, the main conclusions are as follows. 
Long term attractors are nearly circular closed periodic orbits when these exist.
When they don't, which is the case in a nearly homogeneous harmonic core, the only
available attractor is the centre. In that case, one expects significant mass inflow
towards the centre. This process is however {\em self regulating}. When enough mass
has reached the centre, the resulting central concentration destroys  nearly 
constant density of the core and leads to the existence of stable closed loop orbits
(around which dissipative trajectories may settle). 
The central mass that is required to stop further gas inflow is about  
  $0.05 \%$ of the total mass of the galaxy at 20 kpc. 

The above conclusions can also be reached by considering the dynamics of 
a collection of locally interacting ``sticky'' particles. 
In this scenario, particles move 
influenced only by gravity until they come close enough together, when they
collide inelastically. Although such  a procedure might seem at first sight
artificial and somewhat trivial, it can actually be fairly rigorously
justified under certain conditions, if some refinements are introduced (El-Zant 1998) 
--- and is certainly no less justified than a continuum approximation with a perfect
gas equation.
The results indicate that the time-scale for gas inflow is about  a Gyr.
The aforementioned processes leading  to gas accession towards the centre provides a 
self regulating mechanism that does not require the existence of bars. This is important
since it is now thought  (Mulchaey and Regan 1997) that the correlation between activity in the centres of galaxies 
and the existence of stellar bars is weak. 

 When the non-axisymmetric perturbation is rotating (e.g., as in bar of a disk galaxy),
dissipative orbits will follow closed loop orbits when these are stable. Otherwise, 
they will dissipate towards the centre, crossing the variety of resonance 
regions populated by the chaotic replacements of the $x_2$ orbits. During this interval
the motion would exhibit ``transient'' chaotic behaviour (e.g., LL).

In the case when  both a rotating bar  and 
and a  nonrotating non-axisymmetric
perturbation are present,
the potential is inherently time dependent.
Dissipative trajectories can therefore end in strange  attractors on which long lived
chaotic motion can occur. Depending on the parameters of the system 
(dissipation and bar strengths etc.), the radial motion 
could be confined to ``rings'' around the bar or large erratic motions in the cylindrical 
$R$ coordinate can occur. There can also be large excursions in the vertical direction
for orbits starting near the disk plane.

The above effects can have several observable consequences on the evolution of galaxies, 
including the building of bulges and its relation with the shapes of dark matter
haloes, the survival and evolution of bars, star formation activity etc. These are 
discussed in some detail in El-Zant (1998). However, as long as detailed treatment 
of the processes that dominate the evolution of the inter-stellar medium are still
not very well understood, such description will only be schematic. Nevertheless, as we have 
seen here, it is still possible to understand the {\em generic} effects of dissipation 
in galaxy models in fairly simple manner, which can be understood from elementary principles
of dissipative dynamics. It is then perhaps appropriate, when modelling effects in the 
inter-stellar medium, to isolate the essential phenomena and model them in as simple a manner 
as possible --- the most sophisticated hydrodynamic scheme is useless, if fundamental 
assumptions like local thermal equilibrium (essential to that type of description)
are not valid --- by using methods that are easier to implement and interpret.

\newpage

\end{document}